# Gas-phase functionalization of macroscopic carbon nanotube fiber fabrics: reaction control, electrochemical properties and use for flexible supercapacitors

*Daniel Iglesias[a,‡], Evgeny Senokos[b,c,d,‡], Belén Alemán[b], Laura Cabana[b], Cristina Navío[e], Rebeca Marcilla[d], Maurizio Prato[a,f,g], Juan J. Vilatela[b]\*, Silvia Marchesan[a]\*.*

[a] Dipartimento di Scienze Chimiche e Farmaceutiche, Università di Trieste, Via L. Giorgieri 1, 34127 Trieste, Italy

[b] IMDEA Materials Institute, c/ Eric Kandel 2, Getafe 28906, Madrid, Spain

[c] E. T. S. de Ingenieros de Caminos, Universidad Politécnica de Madrid, 28040 Madrid, Spain

[d] IMDEA Energy Institute, Parque Tecnológico de Móstoles, Avda. De la Sagra 3, 28935 Móstoles, Madrid, Spain

[e] IMDEA Nanoscience Institute, Faraday 9, Cantoblanco, 28049 Madrid, Spain

[f] Carbon Nanobiotechnology Laboratory, CIC biomaGUNE, Paseo de Miramón 182, 20009 Donostia-San Sebastian, Spain

[g] Basque Fdn Sci, Ikerbasque, Bilbao 48013, Spain

‡ These authors contributed equally to this work

ABSTRACT

The assembly of aligned carbon nanotubes (CNT) into fibers (CNTF) is a convenient approach to exploit and apply the unique physico-chemical properties of CNTs in many fields. CNT functionalization has been extensively used for their implementation into composites and devices. However, CNTF functionalization is still in its infancy, due to the challenges associated with preservation of CNTF morphology. Here, we report a thorough study of the gas-phase functionalization of CNTF fabrics using ozone that was generated *in situ* from a UV-source. By contrast with liquid-based oxidation methods, this gas-phase approach preserves CNTF morphology, whilst notably increasing its hydrophilicity. The functionalized material is thoroughly characterized by Raman, XPS, TEM and SEM. Its newly acquired hydrophilicity enables CNTF electrochemical characterization in aqueous media, which was not possible for the pristine material. Through comparison of electrochemical measurements in aqueous electrolyte and ionic liquid we decouple the effects of functionalization on pseudocapacitive reactions and quantum capacitance. The functionalized CNTF fabric is successfully used as active material and current collector in all-solid supercapacitor flexible devices with ionic liquid-based polymer electrolyte.

INTRODUCTION

Carbon nanotubes (CNTs) have unique physico-chemical properties with great potential in diverse applications, including catalysis[1–3] and energy storage[4], dye-sensitized solar cells[3], biological use such as drug delivery[5], biosensing[6] and conductive-tissue engineering,[7] and in general functional composites[8–11] and hybrids[12]. To exploit their nanostructure properties on a macroscopic scale they are assembled into hierarchical architectures, such as foams, films or fibers.[13–16] In this regard, fibers of aligned CNTs (CNTF) are particularly attractive since this geometrical arrangement efficiently exploits the anisotropic properties of the building block and thus results in high-performance mechanical properties,[17] high electrical conductivity[18] and superior thermal conductivity relative to copper.[19] These properties coexist with a large porosity, making them ideal electrodes for supercapacitors,[20] batteries,[21] sensors[22] and other devices.[23] Their wide spectrum of application and possibility for large-scale production in industrial facilities make CNTF a nanostructured material of particular technological relevance. Yet, there are many applications that require interfacing CNTF with water, polar solvents, polar precursors (*e.g.* for sol-gel) or metal oxides, in which the hydrophobicity of the highly graphitic CNTs prevent their utilization in pristine form.

Over the last decades different routes have been developed to functionalize CNTs in powder form.[24] Functionalization of CNTs implies the introduction of functional groups on the graphitic surface of the nanomaterial, typically *via* wet-chemical methods. Functional groups play a key role to reduce CNT aggregation propensity, and to confer CNTs with additional properties (*e.g.* light harvesting or linking points for metal oxides). Functionalization is indeed a crucial step to disperse CNTs in homogeneous system and allow tight interaction with components of different chemical nature, although it must be properly fine-tuned not to compromise excessively CNT electronic properties.[25,26]

The case of CNTF filaments and fabrics is particularly challenging since the macroscopic morphology of the anisotropic material must be preserved, making traditional wet-chemical methods that require dispersion and stirring, unsuitable. There is large scope for developing CNTF functionalization protocols. For instance, oxidation in acidic media (*i.e.* $H_2SO_4/HNO_3$) was reported to improve CNTF electromechanical properties for pseudocapacitive redox reactions;[27] alternatively, CNTF cross-linking enhanced their mechanical performance.[28] Other functionalization protocols such as the diazo-coupling or nitrene cycloaddition have been used to cross-link CNTFs.[29,30] However, promising as these reports might be, there is generally poor control of the degree of functionalization across the CNTF material as a consequence of the use of liquid-based derivatization. This is due to the fact that the CNTF is a macroscopic assembly that needs to be pervaded by liquid reagents, and thus, in liquid-phase reactions, CNTF behave substantially different from individual nanostructures that expose their surface to the media, as the case of CNTs. Besides, when we attempted acid-oxidation, 1,3-dipole cycloaddition, or diazo-coupling in liquid-media, the CNTF macroscopic structure was significantly damaged (see ESI). There is clearly the need for developing protocols that accurately control the level of CNTF functionalization, and therefore interfacial processes, yet preserving the morphology of the fibers. Gas-phase reactions are ideal for this purpose, especially those already demonstrated on CNTs in powder form.[31–36] Amongst them, the use of ozone is of particular interest since it can be conveniently generated *in situ* by irradiation of air with a high-energy UV lamp. Indeed, the simplicity of this method has made it a popular route to functionalize different carbon nanostructures.[37–39]

The reaction mechanism of CNT ozone-mediated oxidation is still object of investigation. Yim and Johnson reported a detailed *in silico* study of two possible mechanisms.[40] Both are initiated by formation of a five-membered heterocycle between ozone and a C=C double bond (primary ozonide). The authors analyzed its evolution *via* either classical Criegge's mechanism or formation of an intermediate epoxide. In the first scenario, the heterocycle breaks into a ketone and a superoxide (Criegge's intermediate), with subsequent rupture of the C=C bond of the carbon lattice. This intermediate could evolve to two different lactones with very similar formation energy. These either form a pyran or etch the CNT, requiring very high energy (ca. 39 kcal mol$^{-1}$) and producing either CO or $CO_2$, respectively. In the second scenario, the primary ozonide evolves to an epoxide and produces $O_2$. This step has a much lower energy barrier (7.9 instead of 17.0 kcal mol$^{-1}$) relative to the analogous step in Criegge's mechanism, since it involves no C-C cleavage. The epoxide could then form an ether with an affordable energetic cost at room temperature, thus this second scenario appears more plausible.

In contrast with these models, experimental studies based on XPS data show the presence of many different oxygen-containing functional groups, including COOH, C-OH, C=O and C-O-C.[33,35] Discrepancies between experimental data and theoretical calculations are explained as a consequence of side reactions, occurring between the intermediates/final products and the $O_3$ present in the reaction media.

In this work, we apply UV-generated ozone to functionalize CNTF fabricsand we show the general applicability of the protocol by using CNTFs obtained from different precursors (*i.e.,* toluene or butanol). The treatment is evaluated over increasing reaction times, and the functionalized material stability is assessed over a period of up to three weeks. The degree of functionalization is determined by Raman spectroscopy, transmission electron microscopy (TEM), and X-ray photoelectron spectroscopy (XPS). We discuss the effects of functional groups on tensile and electrical properties, provide extensive electrochemical characterization and assemble flexible all-solid supercapacitors with high energy density. Importantly, comparison of measurements in aqueous electrolyte *versus* ionic liquid enables discrimination of pseudocapacitive effects from an increase in quantum capacitance, both of which produce increments in electric-double layer capacitance and energy density.

RESULTS AND DISCUSSION

CNTFs are synthesized by direct spinning of an aerogel of CNTs that is directly drawn out of the reactor during growth by chemical vapour deposition (CVD).[41] By overlapping multiple individual CNTF filaments it is possible to produce either unidirectional non-woven fabric material (Figure 1a) or strings (Figure 1b). The first format is ideal for electrodes and planar composites, whereas the second is useful as a textile that can be bent, twisted or knotted.

In this work, gas-phase functionalization is carried out on both types of samples, as well as on individual 10 micron-diameter filaments. CNTFs are obtained from either one of two different precursors (*i.e.*, butanol or toluene) to demonstrate the general applicability of the functionalization protocol, and also to study the effect of pre-existing CNTF defects on functionalization level. Samples are functionalized by time-controlled exposure to ozone, which is generated *in situ* with the UV lamp of a UV-Ozone cleaner under ambient conditions.

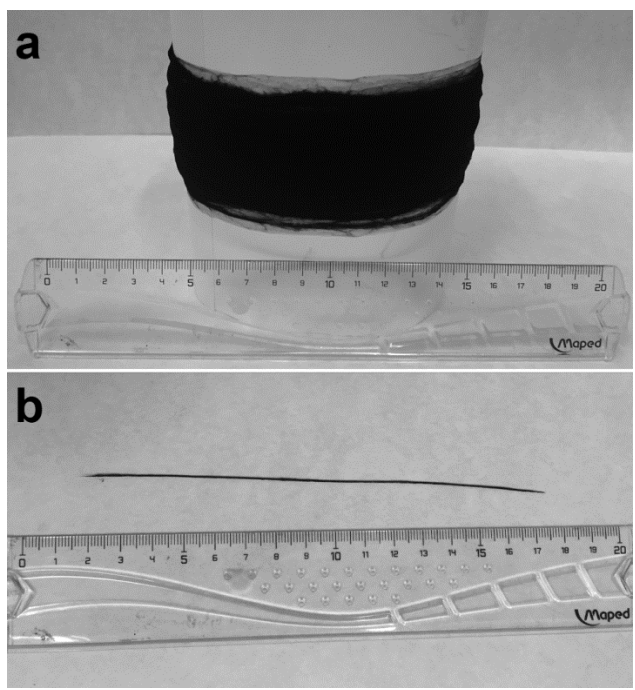

**Figure 1**. Photographs of different CNTF formats. (a) Sample of as-produced CNTF film. (b) String containing 100 individual CNTF filaments.

A first indication of successful CNTF functionalization is the newly acquired hydrophilicity, which locates the CNTF in the aqueous phase of a bi-phasic system composed of toluene-water (Fig.2a). Clearly, water spreads on and penetrates into the functionalized material at a significantly higher rate relative to the pristine CNTF fabric, as shown in Fig.2b. When CNTFs are exposed to ozone for 5 minutes, the water droplet remains fairly spherical upon contact with the CNTF, then gradually infiltrates the porous sample and slowly spreads over it. When CNTFs are exposed to ozone for 2 hours, instant wetting is observed, with more uniform droplet spreading and faster water infiltration. For reference, pristine samples show no evidence of wetting by water on these time scales. Very importantly, the treatment preserved the macroscopic fabric structure unaltered, as confirmed by visual inspection and SEM observation (Figure S2)

TEM observation (Figure 3) confirms partial degradation of CNTs after functionalization. The presence of holes, broken layers and other defect sites along the CNT sidewalls is a consequence of transformation of the C=C bonds to out of plane C-C bonds in the newly-formed functional groups. Such groups are responsible for the hydrophilic behavior observed in Figure 2.

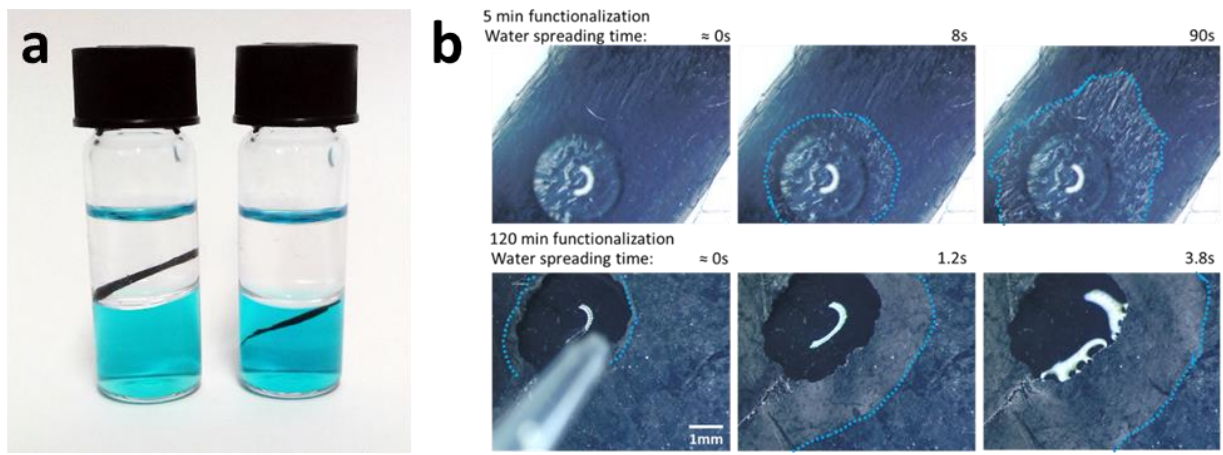

**Figure 2.** Evidence of CNTF hydrophilic nature after functionalization. a) When immersed in a biphasic system composed of toluene-water, pristine CNTFs locate in the organic phase (left), while functionalized CNTFs locate in the aqueous phase (right). Methylene blue was added as a water-soluble dye to visualize more clearly the two phases. b) Optical micrographs monitoring the spreading of a water droplet after contact onto CNTF films functionalized for 5 minutes (top) or 120 minutes (bottom). The dashed lines mark the advance of the liquid front.

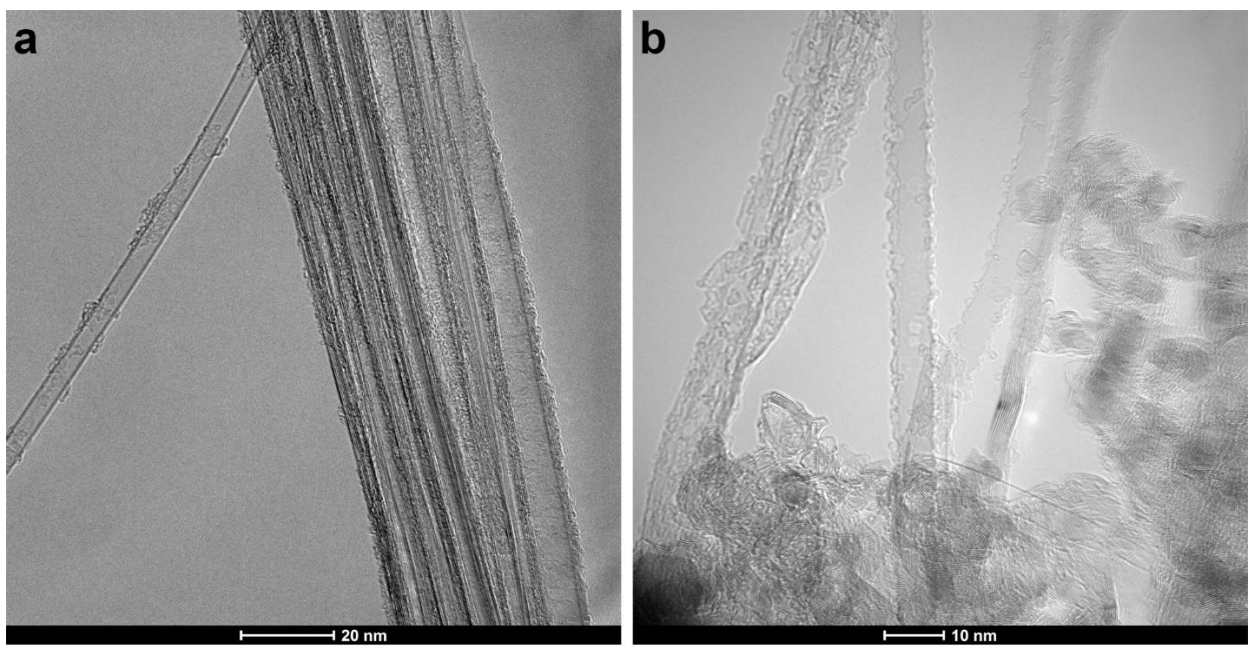

**Figure 3.** TEM images of CNTFs a) before and b) after ozone treatment.

In order to establish a functionalization protocol that not only makes the material hydrophilic, but also provides a degree of surface engineering control in terms of the concentration of functional groups, we analyze in more detail samples subjected to different reaction times. For this purpose, Raman spectroscopy stands as a key technique for the rapid characterization of carbon nanomaterials. For CNTs, the three relevant signals are 1) the second-order D-band due to out-of-plane vibrations from $A_{1g}$ mode (c.a. 1350 cm$^{-1}$), 2) the G-band due to the tangential $E_{2g}$ mode (c.a. 1580 cm$^{-1}$), and 3) 3) the 2D-band (or G') corresponding to the overtone of D-band and resulting from a two-phonon lattice vibrational process (c.a. 2600-2700 cm$^{-1}$).[42] The presence of SWCNTs can be also reflected in appearance of the radial breathing mode (RBM) due to coherent out-of-plane vibration of C-C atoms at 100-500 cm$^{-1}$, whilst substantial fraction of defects in CNTs usually leads to the uprise of a weak shoulder of the G-band at 1615 cm$^{-2}$ corresponding to D* (or G$^+$) band, which is a double resonance feature induced by disorder and defects.[43,44] Raman spectra are shown in Figure 4.

Overall, pristine CNTFs are strongly resonant, with a very low D-band intensity and an asymmetric G-band indicative of few-layer highly graphitized CNTs, with a small fraction of SWCNTs. Raman spectra change rapidly upon exposure to ozone: there is a large drop in absolute intensity due to the loss of resonance, disappearance of RBMs and other few-layer CNT features and the appearance of the D* mode at 1612 cm$^{-1}$, typically activated by defects. To prove the general applicability of this functionalization protocol, CNTFs obtained from either one of two different precursors (*i.e.*, butanol or toluene) are ozone-treated. Raman spectra of pristine CNTFs are similar, with CNTFs obtained from butanol showing a slightly higher level of defects ($I_D$ = 0.13 ± 0.01) relative to CNTFs made from toluene ($I_D$ = 0.09 ± 0.01). This difference is reflected in the relative reactivity of the two CNTF materials, with the former reacting more promptly than the latter, as confirmed by Raman spectra evolution as reaction time was prolonged (Figure 4a *vs.* Figure 4b). In both cases, functionalization is evident.

The relative intensity of the D and G band, $I_D/I_G$, is a convenient parameter to qualitatively evaluate the degree of functionalization of CNTs.[45] Figures 4c-d present $I_D/I_G$ for different reaction times, ranging from 5 minutes to 2 hours. The graph also shows that the effects of functionalization are generally stable over time, a common concern associated with low-temperature UV irradiation or similar treatments.[37, 46] The effect of functionalization is evident already after 5 minutes of exposure to ozone, especially when butanol is used as a precursor. Such trend continues as the treatment is prolonged up to two hours, but at that point the concentration of defects is considered excessive for practical applications. As observed by comparison of the Raman spectra for the samples produced from different precursors, for any given reaction time CNTFs obtained from butanol are more reactive than those obtained from toluene, as shown by their $I_D/I_G$ ratios (compare Fig. 4c with Fig. 4d).

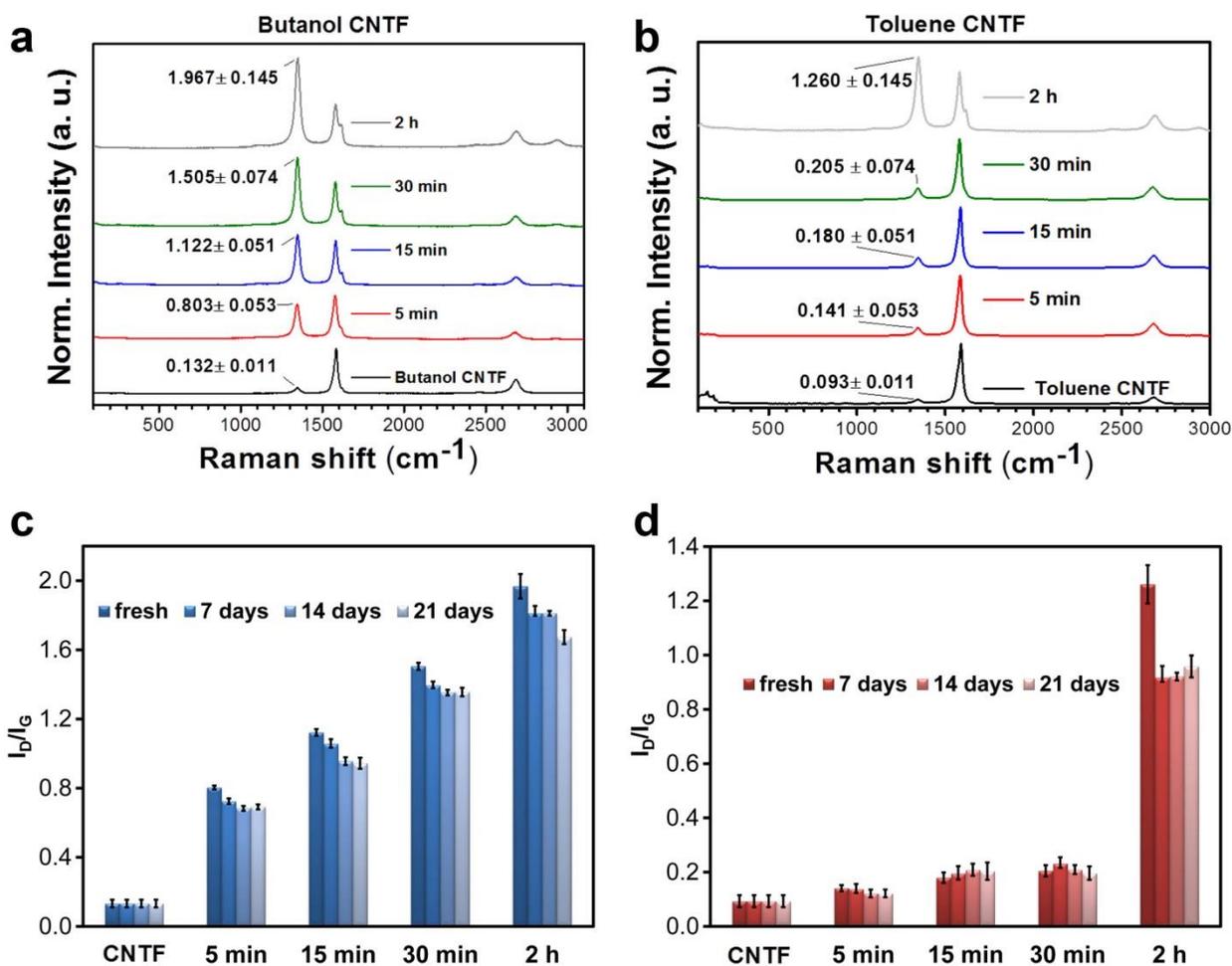

**Figure 4**. Raman characterization for butanol-derived (left) or toluene-derived (right) CNTF upon ozone-treatment. Normalized Raman spectra of pristine and functionalized CNTFs (a, b) and evolution of the $I_D/I_G$ ratio (c, d) upon increase of reaction time from 5 minutes to 2 hours, and its stability over a time up to 3 weeks. Note: Raman spectra are an average of 10 data points.

The incorporation of functional groups on CNTFs is probed with XPS measurements. Figure 6 presents normalised XPS survey spectra for pristine and all ozone treated CNTFs obtained from butanol, showing that the oxygen to carbon relative content, obtained from the peak areas, increases from 0.09 to 0.40. The presence of a small amount of oxygen in the pristine CNTF is attributed mainly to adsorbed molecular species. By contrast, after ozone treatment the observed increase in the O1s region corresponds to the formation with the reaction time of oxygen functional groups bounded to carbon (Table 1).

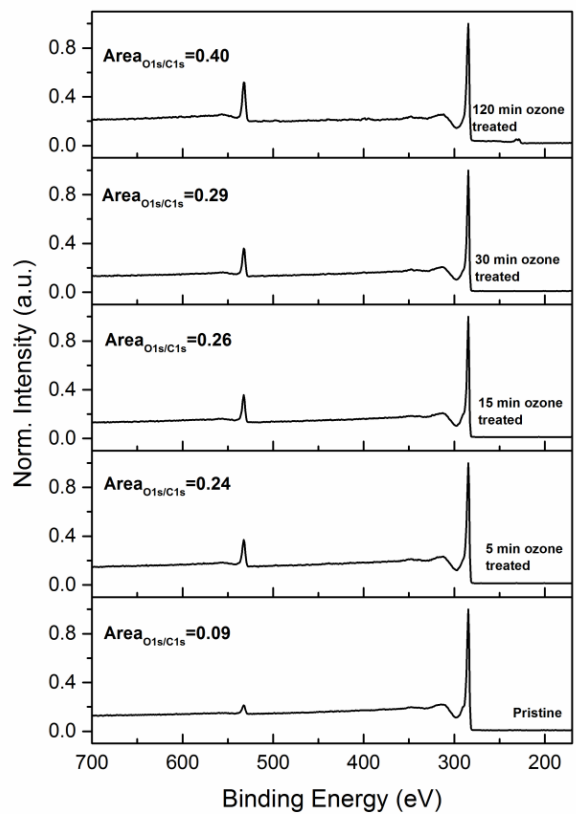

**Figure 5.** Survey XPS spectra of pristine and ozone treated CNTFs produced from butanol with their corresponding relative oxygen to carbon content represented by the peak area ratio of O1s and C1s emissions, which increases with ozone exposure time from 0.09 (pristine) to 0.40 (120 min treatment).

Figure 6 shows XPS spectra for C1s core level of pristine and 120 minute-treated CNTFs with their corresponding fitting components as well as the intermediate ozone treatment exposure times.

The C1s emission of pristine CNTF (Figure 6b) is peaked at 284.5 eV, corresponding to $sp^2$-hybridized C in the CNT structure (C=C). Three smaller intensity components were additionally used to fit the spectrum at 285.1, 286.1 and 290.2 eV. The peak at 285.1 eV originates from $sp^3$ carbon (C-C), whereas the peak at 286.1 eV is associated with oxygen-containing species (C-O).[47,48] Both are associated with carbonaceous particulates and amorphous carbon produced as by-products of the CVD reaction. At the higher binding energy range of 290 eV we can observe π-π* transitions. The presence of this component indicates the high degree of graphitization of the CNTF produced by this method, most likely due to the comparatively high temperature used for CNT growth above 1200°C. Figures 6b-fd show C1s peak deconvolution for all functionalized CNTFs where the increase of the contributions at high binding energies (from 285 eV to 290 eV) compared with $sp^2$ C, together with the gradual loss of the π-π* transition band (Table 1), confirm the introduction of functional groups onto CNTF. A univocal assignment of oxygen-containing components of C1s is very complex, for instance because functional groups affect not only carbon atoms involved in the direct bond but also the next-nearest carbon atoms,[49] nevertheless we present the assignation based on literature[48–50] of C-O at 286.0±0.6 eV, C=O and O-C=O at 289±1 eV. Table 1 also shows how the area ratio of O1s peak with respect to C1s peak (O1s/C1s) increases with ozone treatment exposure time.

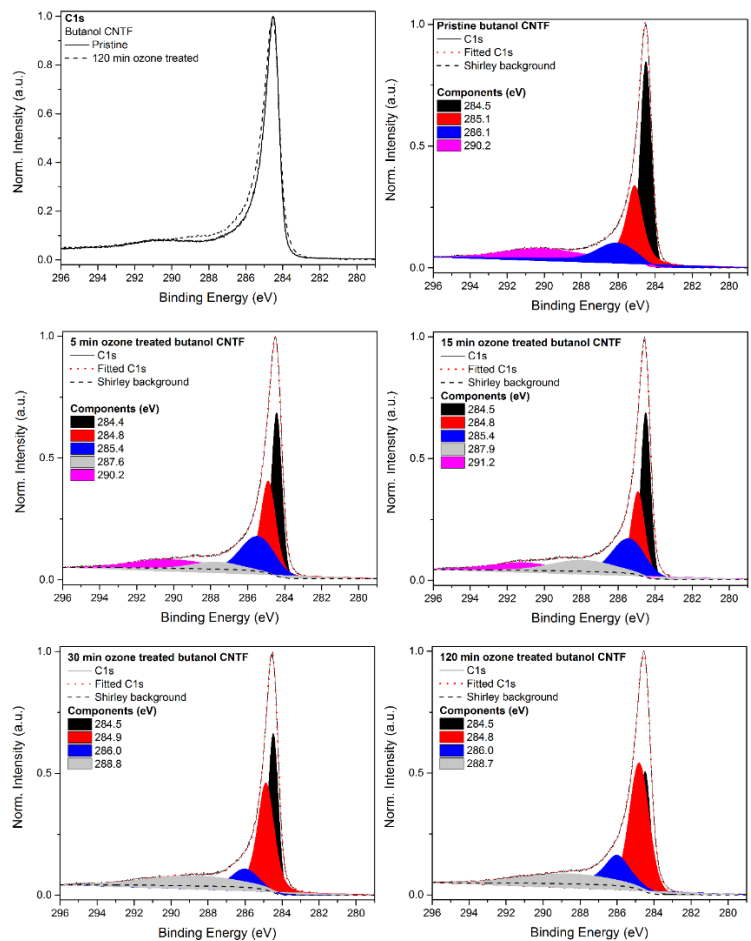

**Figure 6.** XPS C1s emission spectra (a) of pristine and 120 min ozone treated CNTF and (b-f) their deconvolution including the intermediate treatments with components at 284.5±0.1 eV, 285.0±0.2 eV, 286.0±0.6 eV, 289±1 eV and 291±1 eV assigned to $sp^2$-hybridised C (C=C), $sp^3$-hybridised C (C-C), C-O groups, C=O and O-C=O groups, π-π* interaction energy loss, respectively. Increase of ozone treatment time leads to an increase of the contributions at high binding energies (from 285 eV to 290 eV) compared with $sp^2$ C and the gradual loss of the π-π* transition band.

Overall, XPS data are consistent with Raman data, showing a gradual introduction of functional groups at the expense of degrading the graphitic structure of the CNTF. In addition, the comparison of pristine butanol and toluene samples (see ESI) shows the latter to have a higher fraction of graphitic carbon. Defects are the most energetically favorable locations for the formation of functional groups; hence we attribute the differences in functionalization rates detected by Raman to differences in density of pre-existing defects in the two materials.

**Table 1.** Fitting details of C1s XPS spectra and the ratio between oxygen functionalised carbon (C-O, C=O and O-C=O) and $sp^2$ carbon components of pristine and ozone-treated CNTF obtained from butanol. *Reaction time.

|  | C=C | | C-C | | C-O | | C=O O-C=O | | π-π* | | O1s/C1s |
| --- | --- | --- | --- | --- | --- | --- | --- | --- | --- | --- | --- |
|  | Peak (eV) | Area (%) | Peak (eV) | Area (%) | Peak (eV) | Area (%) | Peak (eV) | Area (%) | Peak (eV) | Area (%) |  |
| **Pristine** | 284.5 | 42 | 285.1 | 33 | 286.1 | 12 | 0 | 0 | 290 | 13 | 0.09 |
| **5 min*** | 284.4 | 30 | 284.8 | 30 | 285.4 | 22 | 288 | 6 | 290 | 12 | 0.24 |
| **15 min*** | 284.5 | 30 | 284.9 | 26 | 285.4 | 19 | 288 | 13 | 291 | 12 | 0.26 |
| **30 min*** | 284.5 | 30 | 284.9 | 42 | 286.0 | 9 | 289 | 19 | 0 | 0 | 0.29 |
| **120 min*** | 284.5 | 21 | 284.8 | 45 | 286.0 | 18 | 289 | 16 | 0 | 0 | 0.40 |

In Table 2 we summarize the changes in longitudinal mechanical and electrical properties for butanol samples. We compare individual CNTF filaments, before and after ozone-treatment for 15 minutes. Longer functionalization times reduce significantly sample properties, and thus handling. The net effect of ozone-treatment for shorter time intervals (≤ 15 min.) on tensile properties is to increase stiffness and reduce ductility, while preserving strength (stress-strain curves in ESI). This is a consequence of the removal of surface impurities as well as the introduction of surface functional groups in CNTs taking part in adjacent bundles. The presence of functional groups at the interface between two CNTs without crystallographic registry under shear is expected to improve load transfer; the functional groups protrude normal to the CNT axis and provide corrugation[51] in the otherwise flat potential between turbostratic graphitic layers[52]. A similar behavior has been reported in literature for thermal and plasma oxygen-treated CNT fibers[53,32] as well as in fibers that have been irradiated with an electron beam.[54] Longer treatment produced excessive damage to the samples, making them brittle and difficult to handle for tensile testing.

Longitudinal electrical conductivity of functionalized fibers decreases relative to the pristine material, as is expected for this type of functionalization[35]. After 15 minutes of treatment, samples display a change of conductivity from 11,188 (pristine CNTF) to 5900 S m$^{-1}$. This is a consequence of the derivatization of the conjugated system, with the corresponding binding of otherwise de-localized charge, and also of a reduced contribution from surface adsorbates acting as dopants[55,22].

**Table 2.** Longitudinal mechanical and electrical properties for funtionalized CNTF obtained from butanol.

| Sample | Tensile strength (GPa/SG) | Tensile modulus (GPa/SG) | Electrical conductivity (S/m) |
|---|---|---|---|
| Pristine | 0.46 +/- 0.13 | 21.4 +/- 6.2 | 11.2 * 10$^3$ |
| 15-min functionalized | 0.52 +/- 0.12 | 43.7 +/- 10.7 | 5.9 * 10$^3$ |

Electrochemical properties and all-solid supercapacitor devices

To characterize the electrochemical properties of CNTFs functionalized for different reaction times, samples were tested by CV in 3-electrode cell using 1M KOH electrolyte. Results are shown in Figure 7. CV curves obtained for butanol CNTFs (Figure 7a) indicate that the pristine sample has a low capacitance, which in this case is a consequence of the electrolyte not infiltrating the electrode and instead only interacting with the outer surface of the CNTF. In contrast, the functionalized samples show a considerable increase in capacitance, as well as the presence of a pair of peaks at -0.37 V and -0.43 V (vs Hg/HgO) corresponding to redox reactions of oxygen-containing groups.[56,57] Longer functionalization times result in an increase of these pseudocapacitive peaks and in the overall capacitance of the sample.

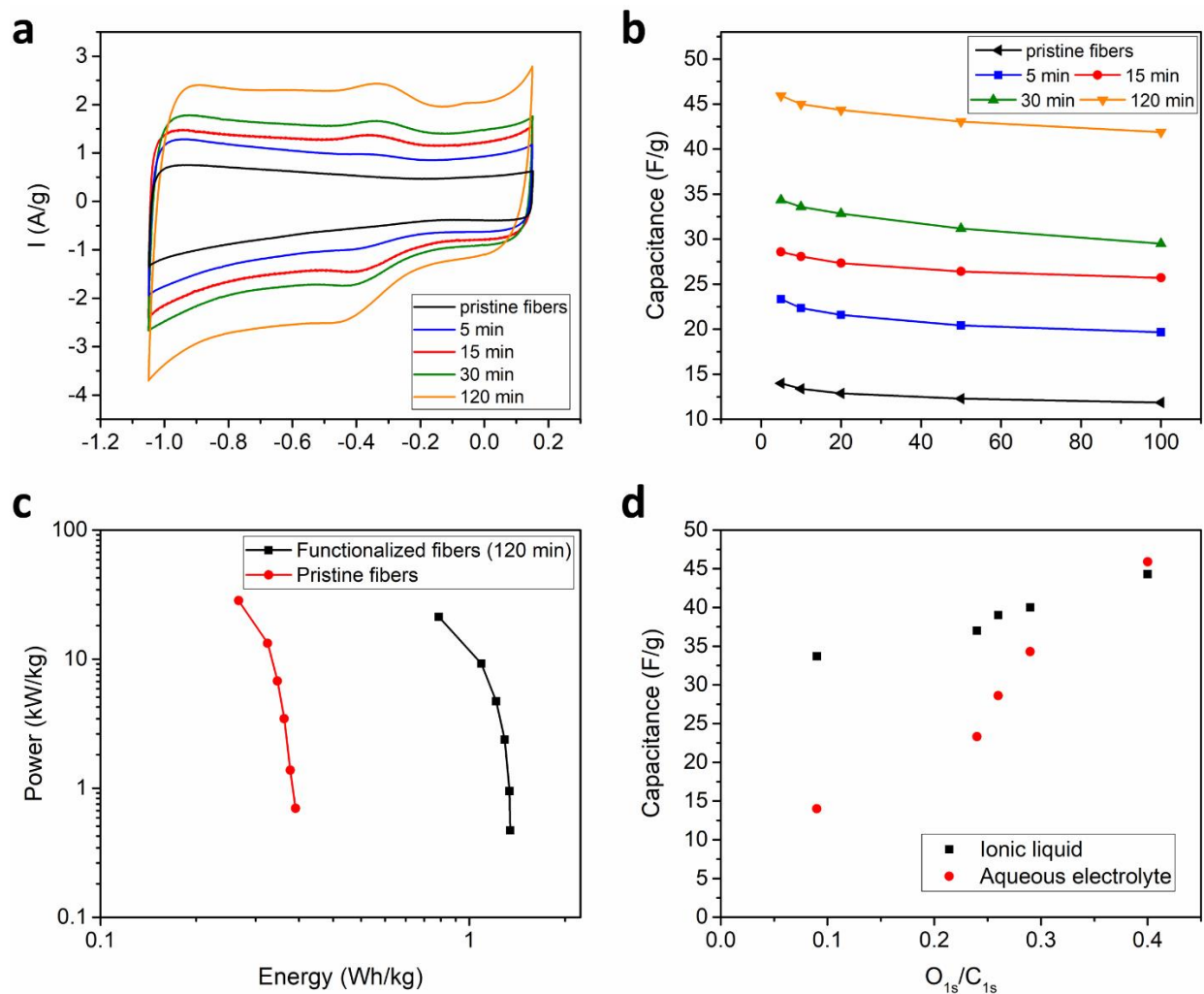

**Figure 7.** (a) CV curves obtained in 1M KOH electrolyte at a scan rate of 50 mV/s and (b) capacitance plots for CNTFs subjected to different functionalization times. (c) Ragone plot comparing electrochemical performance of pristine and 120 min-functionalized CNTFs in symmetric SC device. (d) Specific capacitance against area ratio of O1s respect to C1s extracted from XPS data.

The values of specific capacitance at different scan rates calculated from the area of the CV curves are presented in Figure 7b. A comparison of pristine fibers and samples treated for 5 min, for example, shows a large increase from 14 to 23 Fg$^{-1}$, whereas 2h of exposure leads to 46 Fg$^{-1}$. Functionalised CNTFs retain 90-93% of their capacitance at higher scan rates (Figure 7b), indicating good rate capability due to the combination of easily accessible electrode pores in meso (>10 nm–50 nm) and macro (>50 nm) scale[20,58] with high conductivity and small ions size of the electrolyte. These results are confirmed in symmetric supercapacitor devices assembled in a two-electrode cell. Both charge-discharge and impedance data show an increase in capacitance and no changes in equivalent series resistance from the value for pristine CNTFs around 3 Ω (see ESI). Consequently, the devices with functionalized CNTFs maintain similar values of power at 20-30 kWkg$^{-1}$ while experiencing a large increase in energy density (Figure 7c) from 0.34 to 1.29 Wh kg$^{-1}$ in comparison with pristine CNTFs. These improvements in electrochemical properties are a consequence of the aqueous electrolyte successfully infiltrating the CNTF electrode as it is hydrophilic thanks to functionalization, as well as of the contribution of pseudocapacitive redox reactions to the material total capacitance. Figure 7d shows that specific capacitance in aqueous electrolyte increases monotonically with oxygen to carbon ratio (O1s/C1s) extracted from XPS data.

In addition to the effects mentioned above, functionalization has an effect on the electronic properties of the CNTF, thus on quantum capacitance ($C_Q$) and in the experimentally-determined bulk-electric double-layer EDL capacitance ($C_{EDL}$). These effects can be observed in an ionic liquid (IL), such as 1-butyl-1-methylpyrrolidinium bis(trifluoromethanesulfonyl)imide (PYR$_{14}$TFSI), which produces full wetting and does not undergo redox reactions over a wide potential > 3V (Figure 7d shows that capacitance is not significantly affected by the presence of functional groups in CNTF). A comparison of cyclic voltammograms, now in absence of pseudocapacitive reactions (Figure 8a), shows that functionalized samples lose the characteristic "butterfly" shape of pristine samples and have instead a profile more similar to a bulk 3D carbon.[20] Similarly, EIS measurements give a much weaker dependence of capacitance against electrochemical potential (Figure 8b), and a linear discharge profile for functionalized samples (see ESI). These results demonstrate that functionalization has a large electronic effect on the total capacitance. This is attributed to the introduction of oxygen-containing groups producing new energy states near the Fermi level[59], similar to dopants[60] or charged impurities[61], and which increase quantum capacitance and thus total capacitance. Noting that these capacitive contributions are in series ($\frac{1}{C_{tot}} = \frac{1}{C_Q} + \frac{1}{C_{EDL}}$), the increase in $C_Q$ makes it less accessible, making the electrostatic component of capacitance corresponding to formation of electric-double layer ($C_{EDL}$) more dominant and hence reducing the dependence of total capacitance on electrochemical potential.

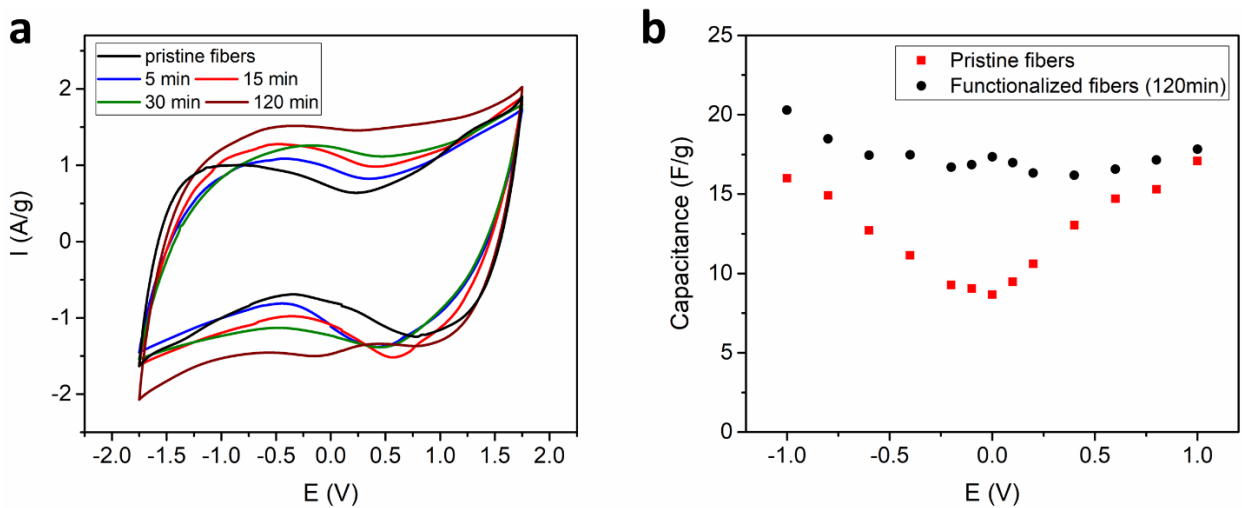

**Figure 8.** (a) CV curves at 50 mV/s and (b) Differential capacitance calculated from electrochemical impedance at 10 mHz obtained for pristine and functionalized CNTF in IL.

Finally, we use functionalized CNTFs to assemble all-solid supercapacitor devices with a polymer electrolyte, following the method described before[62]. Briefly, the process consists of pressing together a sandwich structure of two CNTF electrodes and a membrane of PVDF-HPF and PYR$_{14}$TFSI, without need for a separator. The parameters obtained from CD measurements at 3.5 V are presented in Figure 9. Specific capacitance increases from 27.5 to 43.9 F g$^{-1}$ for functionalized CNTF and energy density reaches a maximum value of 15.4 Wh kg$^{-1}$ at 1 mA cm$^{-2}$ compared to 10.9 Wh kg$^{-1}$ for the pristine material. Considering the electrochemical results in IL discussed above, it is clear that these improvements in capacitance and energy density are not due to pseudocapacitive reactions, rather to an increase in EDL capacitance through a higher quantum capacitance. In Figure 9b (inset) we show an example of a 9 cm$^2$ self-standing all-solid EDLC with functionalized CNT fibers. The device is flexible and can light a red LED in the bent state.

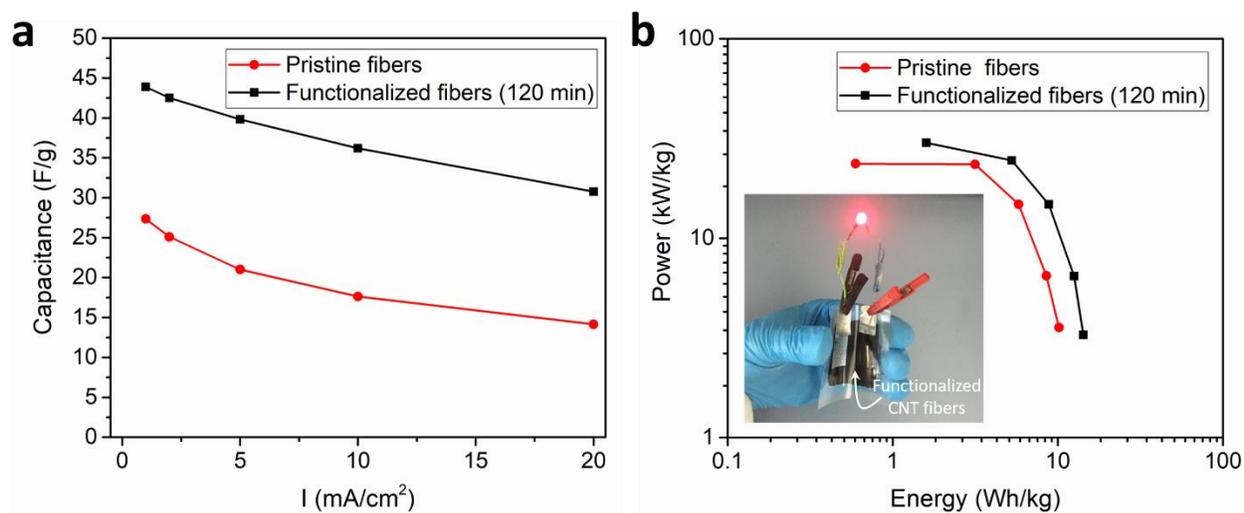

**Figure 9.** (a) Capacitance and (b) Ragone plots demonstrating electrochemical performance of all-solid SCs based on pristine and 120 min-functionalized butanol CNTFs, and self-standing flexible 9 cm$^2$ all solid-state SC based on 120 min-functionalized butanol CNTFs (inset) lighting a red LED in a bent state.

CONCLUSIONS

We describe a very simple method to functionalize CNTFs using UV-generated ozone. We show its general applicability by using CNTFs derived from different precursors. Gas-phase reaction conditions and the absence of purification steps permit the preservation of CNTF morphology at both the macro- and the micro-scale. The degree of CNTF functionalization can be fine-tuned by simply varying the reaction time, as confirmed by Raman and XPS analyses. Importantly, the convenient UV-generated ozone-mediated oxidation of CNTFs renders them hydrophilic and enables use of water-based electrolytes with low cost and high ionic conductivity, which are both attractive for high power density applications. In addition, the ozone treatment increases CNTF specific capacitance, thanks to reversible redox reactions of oxygen-containing functional groups in aqueous electrolyte or to the increase quantum capacitance in aprotic ionic liquids. The newly acquired properties of the CNTF are useful to develop sensors in aqueous systems, or supercapacitors, as shown in an all-solid flexible device.

Overall, this work opens the door to CNTF application in water media, and chemical derivatization of the newly introduced functional groups grants scope of future investigation for the *ad hoc* design of these materials in areas such as sensing and energy storage/harvesting.

EXPERIMENTAL SECTION

CNT fibers were synthesized by the continuous spinning method. It consists in drawing an aerogel of CNTs from the gas phase during their growth by chemical vapor deposition (CVD).[41] The CVD reaction was conducted at 1250 ºC in an $H_2$ atmosphere using ferrocene, thiophene and toluene or butanol as Fe, S and C sources, respectively. The ratio S/C controls the CNT morphology, and it was chosen to produce few-layer multi walled CNT, as reported before.[63] All the samples were spun at a rate of 20 m min$^{-1}$, which corresponds to a draw ratio of 6.3 and leads to preferential alignment of the CNTs parallel to each other and to the fiber axis.[64] The functionalization of CNT fiber was done using a Cleaner ProCleaner™ Plus. The instrument consist in a high energy light (*i.e.*, λ = 185 nm) that produces ozone from atmospheric oxygen (*i.e.*, there is no gas flowing). Initially, CNT ropes were subjected to different reaction times to follow the kinetics of the process (*i.e.*, 5 min, 15 min, 30 min or 2 h).

Optical micrographs were obtained with a S7X10 Olympus microscope. Raman spectroscopy was acquired in an Invia Renishaw microspectrometer (50) equipped with He−Ne laser at 532 nm. All reported spectra are the average of 10 data points and confirm the homogeneity of the materials. Thermogravimetric analysis were recorded in a TGA Q500 (TA instruments). The analyses were performed under air from 100 ºC to 800 ºC using a 10 ºC min$^{-1}$ ramp. TEM images were acquired with a Talos F200X, FEI, operating at 80 KV. TEM samples were prepared by dispersing a small amount of CNT fibers sample in absolute ethanol and sonicated in an ultrasonic power bath. After that, the dispersion was drop cast onto a lacey carbon TEM grid. XPS data were collected in a Phoibos 100 (SPECS GmbH) and SPHERA-U7 (Scienta Omicron GmbH) hemispherical energy analyzers with a monochromatic Al-K$\alpha$ (hv = 1486.71 eV) X-Ray source. To fit XPS spectra Gaussian/Lorentzian peak shapes and Shirley background profile subtraction are used. Mechanical measurements were done with a Textechno Favimat using a gauge length of 20 mm and a strain rate of 0.1/min. Several samples for each condition were tested. Fiber lineal density was determined by weighing a known length of fiber. Fiber modulus was obtained from the intersection of tangent lines to the elastic and plastic regions, determined from linear fitting of the slopes in the derivative of the stress-strain curve. Electircal resistance was determined from current-voltage measurements performed using a microprobe station combined with a 2450 Keithley source-measurement unit, using Indium to create contact with the CNT fibers.

Electrochemical characterization of the carbon material was performed by cyclic voltammetry (CV) using a Biologic VMP multichannel potentiostatic–galvanostatic system and 3 electrode cell configuration with platinum mesh as the counter electrode and Hg/HgO reference electrode. 1M KOH solution and PYR$_{14}$TFSI were used as aqueous and non-aqueous electrolytes, respectively. Scan rates applied ranged from 5 to 100 mV s$^{-1}$ and the voltage window used was 1.2 V. Specific capacitance was obtained by integrating the area under CV curves and normalizing by mass of active material. Electrochemical performance of symmetric SCs was analyzed by galvanostatic charge-discharge with current density range from 1 to 20 mA cm$^{-2}$ and electrochemical impedance spectroscopy with frequency varying from 10 mHz to 200 kHz. SCs were assembled in a two electrode Swagelok® cell using cellulose paper as separator and 1M KOH or PYR$_{14}$TFSI as electrolytes. All-solid-state EDLCs were prepared using polymer electrolyte membranes consisting of PYR$_{14}$TFSI ionic liquid and Poly(vinylidene fluoride-co-hexafluoropropene) (PVDF-*co*-HFP) as both electrolyte and separator. The fabrication of membranes and assembly of free-standing all-solid EDLC devices were described before.[62] Specific capacitance of full SC was calculated from the slope of discharge curve as $C_{cell}$ = I/slope. Specific capacitance of a single electrode in the symmetric device was obtained as $C_s$ = 4 $C_{cell}$. Values of real energy ($E_{real}$) and power ($P_{real}$) densities were calculated by integrating discharge curves of full devices according to the equations:

$$E_{real} = I \int V dt \tag{1}$$

$$P_{real} = \frac{E_{real}}{t_{dis}} \tag{2}$$

ASSOCIATED CONTENT

**Supporting information.**

The Supporting Information is available free of charge on the ACS Publications website. XPS data for pristine CNT fibers produced from different precursors, stress-strain curves, further data from electrochemical measurements of pristine and functionalized samples in different electrolytes and photographs of an all-solid flexible supercapacitors lighting a red LED.

AUTHOR INFORMATION

**Corresponding Author**

* juanjose.vilatela@imdea.org, smarchesan@units.it

**Author Contributions**

The manuscript was written through contributions of all authors. All authors have given approval to the final version of the manuscript.

ACKNOWLEDGMENTS

The authors are grateful to Dr. V. Reguero and Mr. J.C. Fernández-Toribio for assistance with sample preparation and testing. Generous financial support was provided by the European Union Seventh Framework Program under grant agreement 678565 (ERC-STEM), by MINECO (RyC-2014-15115 and MAT2015-64167-C2-1-R), by CAM MAD2D project (S2013/MIT-3007), and by the Italian Ministry of University and Research MIUR under the PRIN 2015 program (Grant n. 2015TWP83Z). Part of this work was supported by a STSM Grant from COST Action CA15107 (MultiComp).

TOC

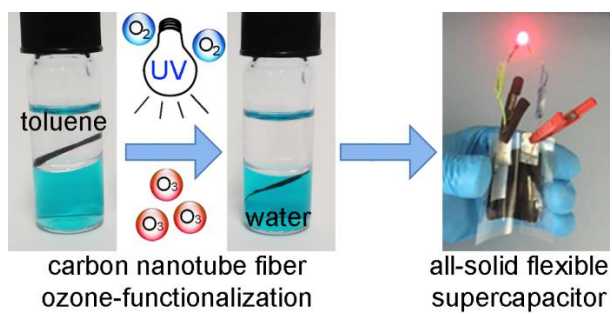

The enhanced hydrophilicity of functionalized carbon nanotube fibers (CNTFs) significantly extends their application in supercapacitors. Well-studied electrochemical behavior of CNTFs demonstrates a promising potential as multifunctional electrodes for all-solid and flexible electric double-layer capacitors (EDLCs). However, despite good performance of the material achieved in ionic liquids, electrochemical application of highly hydrophobic CNTFs remains limited in aqueous electrolytes due to poor infiltration of water solutions into CNTF porous structure. A convenient UV-generated ozone-mediated oxidation of CNTFs renders them hydrophilic and enables use of water-based electrolytes with low cost and high ionic conductivity, which are both attractive for high power density applications. In addition, the ozone treatment increases CNTF specific capacitance, thanks to reversible redox reactions of oxygen-containing functional groups.